\documentstyle[prd,tighten,aps]{revtex}

\newcommand{\be}{\begin{equation}}
\newcommand{\ee}{\end{equation}}
\newcommand{\bn}{\begin{eqnarray}}
\newcommand{\en}{\end{eqnarray}}

\begin{document} 
\draft 
\twocolumn[\hsize\textwidth\columnwidth\hsize\csname
@twocolumnfalse\endcsname

%\begin{document}
\title{Interference phenomena, chiral bosons and Lorentz invariance}
\author{E. M. C. Abreu and A. de Souza Dutra}
\address{Departamento de F\'\i sica e Qu\'\i mica, Universidade Estadual Paulista, \\
Av. Ariberto Pereira da Cunha 333, Guaratinguet\'a, 12516-410, S\~ao Paulo,
SP, Brazil, \\
E-mail: everton@feg.unesp.br and dutra@feg.unesp.br}
\date{\today}
%\draft
\maketitle

\begin{abstract}
\noindent We have studied the theory of gauged chiral bosons and
proposed a general theory, a master action, that encompasses different kinds of gauge
field couplings in chiral bosonized theories with first-class chiral
constraints. We have fused opposite chiral aspects of this master action using the soldering
formalism and applied
the final action to several well known models. The Lorentz rotation
permitted us to fix conditions on the parameters of this general theory in
order to preserve the relativistic invariance. 
We also have established some conditions on
the arbitrary parameter concerned in a chiral Schwinger model with a
generalized constraint, investigating both covariance and Lorentz invariance.
The results obtained supplements the one that shows the soldering formalism as a new method of mass generation.
\end{abstract}
\pacs{11.10.Ef, 12.10.Gq, 04.65.+e}

%end wide text
\vskip2pc]

\newpage

\section{Introduction}

The research in chiral bosonization has begun many years back with the
seminal paper of W. Siegel \cite{siegel}. Floreanini-Jackiw have offered some different solutions to the problem of a single self-dual field \cite{fj}. The study of chiral bosons has blossomed thanks to the advances in superstring compactification \cite{ghmr} and in the construction of interesting theoretical models, such as the Thirring model \cite{grs}. They also play an
important role in the studies of the quantum Hall effect \cite{ws}.  The introduction of a soliton field as a charge-creating field obeying one additional equation of motion leads to a bosonization rule \cite{ggkrs}.
In the course of the analysis of
the chiral boson properties, one natural step is to couple them to Abelian
and non-Abelian gauge fields \cite{bbs,bgp} in order to study the
correspondent anomalies, or to provide an alternative approach to chiral
models in two dimensions \cite{harada}. Gates and Siegel showed how to
construct general interacting actions for chiral bosons, including the
supersymmetric and the non-Abelian cases \cite{gs}. They used this
construction to obtain the righton-lefton interaction by carrying out the
path integral quantization in a generalized Thirring model. Another couplings of the chiral bosons with supersymmetry and gravity can be found in references \cite{bellucci}.
In an alternative way, Stone \cite{ms} has shown that the method of
coadjoint orbit, when applied to a representation of a group associated with
a single affine Kac-Moody algebra, generates an action for the chiral WZW
model \cite{wzw}, a non-Abelian generalization of the Floreanini and Jackiw
(FJ) model. The formalism introduced by Stone could be interpreted recently as a new method of mass generation \cite{abw}.

In the context of chiral theories in two dimensions, Harada has shown \cite
{harada} how to obtain a consistent coupling of FJ chiral bosons with a $U(1)
$ gauge field, starting from the chiral Schwinger model and discarding the
right-handed degrees of freedom by means of a projection in phase space
implemented by the chiral constraint $\pi_{\phi}=\phi^{\prime}$. Later on,
it has been observed that, starting with a chiral Schwinger model  
of a given chirality it is possible to couple chiral bosons to $U(1)$ gauge 
fields in two Lorentz invariant ways, using different chiral constraints  
\cite{dd1,dd2}. 
The theory proposed was shown to be
equivalent to a specific coupling of Siegel's chiral bosons with $U(1)$
gauge fields which is symmetric under chirality-preserving gauge
transformations.

In \cite{bgp}, Bellucci, Golterman and Petcher introduced an $O(N)$
generalization of Siegel's model for chiral bosons coupled to Abelian and
non-Abelian gauge fields. The physical spectrum of the resulting Abelian
theory is that of a (massless) chiral boson and a free massive scalar field.
Bazeia \cite{bazeia} showed that the Bellucci {\it et al} model is
equivalent, at the classical level, to the gauged FJ chiral boson found by
Harada.

In this work we have proposed a
general action to describe the gauge coupling in different chiral
bosonization schemes. We used the soldering formalism proposed by Stone \cite{ms} to fuse opposite chiralities producing a final
action which was used to apply to various chiral theories with new outcomes.
We have also analyzed the problem of the Lorentz invariance of this general
model. In \cite{dd1}, a bosonized form of the chiral Schwinger model with a
generalized constraint was analyzed using the Lorentz rotation to fix a
general parameter. The soldering formalism permitted us to fix conditions on this parameter to
obtain manifest covariance or, in another case, Lorentz invariance of this 
self-dual action.

We have organized the paper in the following way: in section 2 we have tried
to make a self-consistent review of the soldering formalism
and used the well known Siegel's theory as an example to clarify the
interference concepts. In section 3 we have introduced a general action, a master action,
which encompass different gauged self-dual actions with first-class
constraints. The technique of soldering has been applied and the final
soldered action was used in several models with new results. In section 4 we
have employed the Lorentz rotation to fix the value of the parameters in
order to guarantee the relativistic invariance of the theory. In section 5
the interference effect has been analyzed using the chiral Schwinger model
with a generalized constraint. The parameter dependence were placed in the
light of the manifest covariance and of the Lorentz invariance of the
soldered action. The conclusions are depicted in section 6.

\section{Review of the soldering formalism}

In this section we will follow basically the references \cite{w2,adw} to
make a short, but at the same time self-consistent review of the method of
soldering two opposite chiral versions of a theory. 

The soldering formalism gives an useful bosonization scheme for
Weyl fermions, since a level one representation of LU(N) has an
interpretation as the Hilbert space for a free chiral fermion \cite{ps}.
However, only Weyl fermions can be analyzed in this way, since a $2D$
conformally invariant QFT has separated right and left current algebras. In
other words, it is trivial to make a (free) Dirac fermion from two (free)
Weyl fermions with opposite chiralities. The action is just the sum of two
Weyl fermion actions. It seems, however, non-trivial to get the action of
the WZW model from two chiral boson actions of opposite ``chiralities'',
because it is not the sum of the two.

To solve this problem, Stone \cite{ms} introduced the idea of soldering the
two chiral scalars by introducing a non-dynamical gauge field\footnote{In
\cite{kh2}, Harada have proposed a physical interpretation for these
soldering fields.} to remove the
degree of freedom that obstructs the vector gauge invariance \cite{w2}. This
is connected, as we said above, to the necessity that one must have more
than the direct sum of two fermions representations of the Kac-Moody algebra
to describe a Dirac fermion. In another way we can say that the equality for
the weights in the two representations is physically connected with the
necessity to abandon one of the two separate chiral symmetries, and accept
that the vector gauge symmetry should be maintained. This is the main
motivation for the introduction of the soldering field which makes possible
the fusion of dualities in all space-time dimensions. Besides, being just an
auxiliary field, it may posteriorly be eliminated in favor of the physically
relevant quantities. This restriction will force the two independent chiral
representations to belong to the same multiplet, effectively soldering them
together. We will see below, in a precise way, more details about the
physical significance of the soldering field.

It is worth to mention that the soldering procedure has a typical quantum
mechanical nature, with no classical analogue. It has no sense to sum two
classical actions, that although describing opposite aspects of some
(duality) symmetry, would depend on the same field. On the other hand, the
direct sum of duality symmetric actions depending on different fields would
not give anything new. It is the soldering process that leads to a new and
non trivial result.

In \cite{adw}, the authors have promoted the soldering the two (Siegel)
invariant representations of opposite chiralities. The symmetry content of
each theory is well described by the Siegel algebra, a truncate
diffeomorphism, that disappear at the quantum level. The resulting action
is invariant under the full diffeomorphism group, which is not
a mere sum of two Siegel symmetries. As we will see later, the result can 
{\it also} be seen as a scalar field immersed in a gravitational background.

Recently, there has been a great deal of interest in soldering together distinct
manifestations of duality. The procedure leads to new physical results 
including quantum contributions. For instance, these results provided the idea of an
interference effect. However, this ``wave'' interpretation is not new. E.
Witten, in \cite{wzw}, associated the fields depending on only one chirality
to left-moving or right-moving waves as being the $\gamma ^{5}$ eigenstates.

One of us, with collaborators \cite{abw}, has promoted the interference of
two chiral Schwinger models with opposite chiralities. As a result it was
obtained a new method of mass generation. The Bose symmetry fixed the
Jackiw-Rajaraman parameter ($a=1$) \cite{jr} so that in the spectrum only
massless harmonic excitations have survived. The soldered action represents a vector 
Schwinger model which has a
massive particle spectrum. This behavior characterizes a constructive
interference with the arising of a mass term that is typical of the
right-left quantum interference \cite{jackiw}\footnote{%
The extension of this case to the four dimensional one was performed in \cite
{bw2}.}. 

In terms of degrees of freedom we can say that each (chiral) action
contributes with ``one half" degree of freedom of opposite signals. Hence, the
soldered action have one degree of freedom.  By the way, in the reference 
\cite{dutra}, it was shown that the direct sum of two CSM with opposite 
chiralities is, in fact, equivalent to a sum of a vector Schwinger model (VSM)
and an axial Schwinger model (ASM), so, getting a different number of degrees
of freedom from a sum of isolated CSM.

It was shown lately \cite{aw}, that in the soldering process of two Siegel's 
\cite{siegel} modes (lefton and righton) coupled to a gauge field \cite{gs},
this gauge field has decoupled from the physical field. The final action
describes a non-mover field (a noton) at the classical level. The noton
acquires dynamics upon quantization. This field was introduced by Hull \cite
{hull} to cancel out the Siegel anomaly. It carries a representation of the
full diffeomorphism group, while its chiral components carry the
representation of the chiral diffeomorphism.

In the $3D$ case, the soldering mechanism was used to show the result of
fusing together two topologically massive modes generated by the
bosonization of two massive Thirring models with opposite mass signatures in
the long wave-length limit. The bosonized modes, which are described by self
and anti-self dual Chern-Simons models \cite{tpn,dj}, were then soldered into
the two massive modes of the $3D$ Proca model \cite{bw}. In the $4D$ case,
the soldering mechanism produced an explicitly dual and covariant action as
the result of the interference between two Schwarz-Sen \cite{ss} actions
displaying opposite aspects of the electromagnetic duality \cite{bw}.

Wotzasek \cite{wotzasek} has obtained the field theoretical analogue of the
``quantum destructive interference'' phenomenon, by coupling the non-Abelian
chiral scalars to appropriately truncated metric fields, known as chiral WZW
models, or non-Abelian Siegel models \cite{wzw}. In fact, this effective
action does not contain either right or left movers, but can be identified
with the non-Abelian generalization of the bosonic non-mover action proposed
by Hull.

In a recent work \cite{ainw}, it was analyzed the restrictions posed by the
soldering formalism over a new regularization class that extends the
classification of the regularization ambiguity of $2D$ fermionic determinant
from three to a four-constraint class. This analysis results from the
interference effects between right and left movers, producing a massive
vectorial photon that constrains the regularization parameter to 
this four-constraints class. In other words, the new Faddeevian class of
chiral bosons proposed by Mitra \cite{mitra} has interfered constructively
to produce a massive vectorial mode.

The basic idea of the soldering procedure is to raise a global Noether
symmetry of the self and anti-self dual constituents into a local one, but
for an effective composite system, consisting of the dual components and an
interference term. The objective in \cite{w2} is to systemize the procedure
like an algorithm and, consequently, to define the soldered action.

An iterative Noether procedure was adopted in \cite{w2} to lift the global
symmetries. Therefore, assume that the symmetries in question are being
described by the local actions $S_{\pm}(\phi_{\pm}^\eta)$, invariant under a
global multi-parametric transformation

\begin{equation}  \label{ii10}
\delta \phi_{\pm}^\eta = \alpha^\eta\;\;,
\end{equation}
where $\eta$ represents the tensorial character of the basic fields in the
dual actions $S_{\pm}$ and, for notational simplicity, will be dropped from
now on. As it is well known, we can write,

\begin{equation}
\delta S_{\pm}\,=\,J^{\pm}\,\partial_{\pm}\,\alpha\;\;,
\end{equation}
where $J^{\pm}$ are the Noether currents.

Now, under local transformations these actions will not remain invariant,
and Noether counter-terms become necessary to reestablish the invariance,
along with appropriate auxiliary fields $B^{(N)}$, the so-called soldering
fields which has no dynamics. Nevertheless we can say that $B^{(N)}$ is an
auxiliary field which makes a wider range of gauge-fixing conditions
available \cite{kh2}. In this way, the $N$-action can be written as,

\begin{equation}  \label{ii20}
S_{\pm}(\phi_{\pm})^{(0)}\rightarrow S_{\pm}(\phi_{\pm})^{(N)}=
S_{\pm}(\phi_{\pm})^{(N-1)}- B^{(N)} J_{\pm}^{(N)}\;\;.
\end{equation}
Here $J_{\pm}^{(N)}$ are the $N-$iteration Noether currents. For the self
and anti-self dual systems we have in mind that this iterative gauging
procedure is (intentionally) constructed not to produce invariant actions
for any finite number of steps. However, if after N repetitions, the non
invariant piece end up being only dependent on the gauging parameters, but
not on the original fields, there will exist the possibility of mutual
cancelation if both self and anti-self gauged systems are put together.
Then, suppose that after N repetitions we arrive at the following
simultaneous conditions,

\begin{eqnarray}  \label{ii30}
\delta S_{\pm}(\phi_{\pm})^{(N)} \neq 0  \nonumber \\
\delta S_{B}(\phi_{\pm})=0\;\;,
\end{eqnarray}
with $S_B$ being the so-called soldered action 
\begin{equation}  \label{ii40}
S_{B}(\phi_{\pm})=S_{+}^{(N)}(\phi_{+}) + S_{-}^{(N)}(\phi_{-})+ %
\mbox{Contact Terms}\;\;,
\end{equation}
where the Contact Terms are generally quadratic functions of the soldering
fields. 
Then we can immediately identify the (soldering) interference term as, 
\begin{equation}  \label{ii50}
S_{int}=\mbox{Contact Terms}-\sum_{N}B^{(N)} J_{\pm}^{(N)}\;\;.
\end{equation}
Incidentally, these auxiliary fields $B^{(N)}$ may be eliminated, for instance,
through its equations of motion, from the resulting effective action, in
favor of the physically relevant degrees of freedom. It is important to
notice that after the elimination of the soldering fields, the resulting
effective action will not depend on either self or anti-self dual fields $%
\phi_{\pm}$ but only in some collective field, say $\Phi$, defined in terms
of the original ones in a (Noether) invariant way

\begin{equation}
S_{B}(\phi _{\pm })\rightarrow S_{eff}(\Phi )\;\;.  \label{ii60}
\end{equation}
Analyzing in terms of the classical degrees of freedom, it is obvious that we have now a
bigger theory. Once such effective action has been established, the physical
consequences of the soldering are readily obtained by simple inspection.
This will progressively be clarified in the specific application to be given
next.

In order to
present an example, we will analyze the Siegel chiral actions in the light
of the interference phenomenon
\footnote{%
We will follow the steps given in \cite{adw}.}. First of all, we have to
describe the light-front variables used in this paper as, 
\begin{eqnarray}
x_{\pm}&=&{\frac{1 }{\sqrt{2}}} (x_0\pm x_1)\;\;,  \nonumber \\
\partial_{\pm}&=&{\frac{1 }{\sqrt{2}}}(\partial_0\,\pm\,\partial_1)\;\;, 
\nonumber \\
A_{\pm}&=&{\frac{1 }{\sqrt{2}}}(A_0\,\pm\,A_1)\;\;,
\end{eqnarray}
and now we can work out our example.

\subsection{An example: the Siegel action}

The original classical Lagrangian density for a chiral scalar field as introduced by Siegel \cite{siegel} for a left moving scalar (a lefton) is \cite{fs}
%The Siegel action for a left-moving chiral boson, a lefton (which will be
%explained below), is 
\begin{eqnarray}  \label{03}
{\cal L}_0^{(+)} &=& \partial_+\varphi\partial_-\varphi\,+\,\lambda_{++}
\partial_-\varphi\partial_-\varphi \nonumber \\
&=& {1 \over 2}\,\sqrt{g}\,g^{\alpha\beta}\,\partial_{\alpha}\varphi\,\partial_{\beta}\varphi\;\;,
\end{eqnarray}
where the metric is given by
\bn
g^{++}\,=\,0\:\:\:\:&,&\:\:\:\:g^{+-}\,=\,1 \nonumber \\
g^{--}\,&=&\,2\,\lambda_{++}\;\;.
\en

The Lagrangian (\ref{03}) is invariant under Siegel gauge symmetry which is an invariance under the combined coordinate transformation and a Weyl rescaling of the form
\bn \label{11}
x_- \rightarrow \tilde{x}_-\,&=&\,x_-\,-\,\epsilon\,(x_+,x_-) \nonumber \\
\delta_w\,g_{\alpha\beta}\,&=&\,-\,g_{\alpha\beta}\,\partial_-\,\epsilon^-\;\;,
\en
where $\epsilon^{\pm}=\epsilon^{\pm}\,(x_{\pm})$.

The fields $\varphi$ and $\lambda_{++}$ transform under (\ref{11}) as follows:
\bn \label{12}
\delta\,\varphi\,&=&\,\epsilon^-\,\partial_-\varphi\;\;, \nonumber \\
\delta\,\lambda_{++}\,&=&\,-\,\partial_+\,\epsilon\,+\,\epsilon\,\partial_+\,\lambda_{++}\,-\,\lambda_{++}\,\partial_+\,\epsilon^-\;\;.
\en
In addition (\ref{03}) is invariant under the global axial transformation
\be
\varphi \rightarrow \tilde{\varphi}\,=\,\varphi\,+\,\bar{\varphi}\;\;,
\ee
where we have currents associated with this axial symmetry.  It is beyond our work to write explicitly these axial currents as well the conserved vector current.  These objects can be found in literature (see \cite{fs} for example).

The symmetry (\ref{12}) describes a lefton. This is the main difference between a
lefton (righton) and a left-moving (right-moving) FJ particle. The first is
provided with symmetry and dynamics, while the second is responsible only
for the dynamics of the theory. We can also say that the lefton (or righton)
carries the anomaly of the system \cite{aw} (the well known Siegel anomaly), since it is relative to the
symmetry of the theory.

Similarly, one can gauge the semi-local affine symmetry 
\begin{eqnarray}  \label{06}
\delta\,\varphi\,&=&\,\epsilon^+\partial_+\varphi  \nonumber \\
\delta\,\lambda_{--}\,&=&\,-\partial_-\epsilon^+
+\epsilon^+\partial_+\,\lambda_{--}-\lambda_{--}\partial_+\epsilon^+\,.
\end{eqnarray}
to obtain the righton.  Next we will promote the fusion of the righton and the lefton obtaining the final soldered action.

\subsection{The soldering procedure}

In fact, if we construct the righton and lefton chiral boson actions as 
\begin{equation}  \label{08}
{\cal L}_0^{(\pm)} = {\frac{1}{2}} J_{\pm}(\varphi)\partial_{\mp}\varphi
\end{equation}

\noindent with 
\begin{equation}  \label{09}
J_{\pm}(\varphi)=2\left(\partial_{\pm}\varphi
+\,\lambda_{\pm\pm}\partial_{\mp}\varphi\right)\;\;,
\end{equation}
\noindent it is easy to verify that these models are indeed invariant under
Siegel's transformations (\ref{12}) and (\ref{06}), using that 
\begin{equation}  \label{010}
\delta J_{\pm}= \epsilon_{\pm}\partial_{\mp}J_{\pm}\;\;.
\end{equation}

\noindent It is worth mentioning at this point that Siegel's actions for
leftons and rightons can be seen as the action for a scalar field immersed
in a gravitational background whose metric is appropriately truncated. In
this sense, Siegel symmetry for each chirality can be seen as a truncation
of the reparametrization symmetry existing for the scalar field action. We
should mention that the Noether current $J_+$ defined above is in fact the
non vanishing component of the left chiral current $J_+ = J_{(L)}^-$, while $%
J_-$ is the non vanishing component of the right chiral current $J_- =
J_{(R)}^+$, with the left and right currents being defined in terms of the
axial and vector currents as 
\begin{eqnarray}  \label{333}
J_\mu^{(L)}=J_\mu^{(A)}+J_\mu^{(V)}\,,  \nonumber \\
J_\mu^{(R)}=J_\mu^{(A)}-J_\mu^{(V)}\,.
\end{eqnarray}

Let us next consider the question of the vector gauge symmetry. We can use
the iterative Noether procedure described above to gauge the global U(1)
symmetry 
\begin{eqnarray}  \label{20}
\delta\varphi &=& \alpha\,,  \nonumber \\
\delta\lambda_{++}\, &=& 0
\end{eqnarray}

\noindent possessed by Siegel's model (\ref{03}). Under the action of the
group of transformations (\ref{20}), written now as a local parameter, the
action (\ref{03}) changes as 
\begin{equation}  \label{30}
\delta {\cal L}_0^{(+)} = \partial_-\alpha J_+
\end{equation}

\noindent with the Noether current $J_+=J_+(\varphi)$ being given as in (\ref
{09}). To cancel out this piece, we introduce the soldering field $B_-$
coupled to the Noether current, redefining the original Siegel's Lagrangian
density as 
\begin{equation}  \label{50}
{\cal L}_0^{(+)}\rightarrow {\cal L}_1^{(+)}={\cal L}_0^{(+)} +B_- J_+ \,,
\end{equation}

\noindent where the variation of the gauge field is defined conveniently as 
\begin{equation}  \label{60}
\delta B_-=-\partial_-\alpha\,.
\end{equation}

\noindent As the variation of ${\cal L}_1^{(+)}$ does not vanish modulo
total derivatives, we introduce a further modification as 
\begin{equation}  \label{70}
{\cal L}_1^{(+)}\rightarrow {\cal L}_2^{(+)}={\cal L}_1^{(+)}+
\lambda_{++}B_-^2
\end{equation}
\noindent whose variation gives 
\begin{equation}  \label{80}
\delta{\cal L}_2^{(+)} = 2 B_-\partial_+\alpha\,.
\end{equation}

\noindent This piece cannot be canceled by a Noether counter-term,
so that a gauge invariant action for $\varphi$ and $B_-$ does not exist, at
least with the introduction of only one gauge field. We observe, however,
that this action has the virtue of having a variation dependent only on $B_-$
and $\alpha$, and not on $\varphi$. Expression (\ref{80}) is a reflection of
the standard anomaly\footnote{%
The soldering analysis of the anomaly has been depicted in \cite{ad}.} that
is intimately connected with the chiral properties of $\varphi$.

Now, if the same gauging procedure is followed for an Siegel boson of
opposite chirality, say 
\begin{equation}  \label{90}
{\cal L}_0^{(-)} = \partial_+\rho\partial_-\rho + \lambda_{--}
\partial_+\rho\partial_+\rho
\end{equation}

\noindent subject to 
\begin{eqnarray}  \label{100}
\delta\rho &=& \alpha\,,  \nonumber \\
\delta\lambda_{--} &=& 0\,,  \nonumber \\
\delta B_+&=& -\partial_+\alpha\,,
\end{eqnarray}

\noindent then one finds that the sum of the right and left gauged actions $%
{\cal L}_2^{(+)}+{\cal L}_2^{(-)}$ can be made gauge invariant if a contact
term of the form 
\begin{equation}  \label{110}
{\cal L}_C = 2 B_+ B_-
\end{equation}

\noindent is introduced. One can check that indeed the complete gauged
Lagrangian 
\begin{eqnarray}  \label{120}
{\cal L}_{TOT} &=& \partial_+\varphi\partial_-\varphi + \lambda_{++}
\partial_-\varphi\partial_-\varphi + \partial_+\rho\partial_-\rho  \nonumber \\
& &+\,\lambda_{--} \partial_+\rho\partial_+\rho + B_+J_-(\rho) + B_- J_+(\varphi) \nonumber \\
& &+\lambda_{--}B_+^2 +\,\lambda_{++}B_-^2 +2 B_- \, B_+
\end{eqnarray}

\noindent with $J_{\pm}$ defined in Eq.(\ref{09}) above, is invariant under
the set of transformations (\ref{20}), (\ref{60}) and (\ref{100}). For
completeness, we note that Lagrangian (\ref{120}) can also be written in the
form

\begin{eqnarray}  \label{121}
& &{\cal L}_{TOT} = D_+\varphi D_-\varphi + \lambda_{++} D_-\varphi D_-\varphi
\nonumber \\
& &+\,D_+\rho D_-\rho + \lambda_{--} D_+\rho D_+\rho
+\left(\varphi-\rho\right)E\,,
\end{eqnarray}

\noindent modulo total derivatives. In the above expression, we have
introduced the covariant derivatives $D_{\pm}\varphi=
\partial_\pm\varphi+B_\pm$, with a similar expression for $D_\pm\rho$, and $%
E\equiv\partial_+B_--\partial_-B_+$. In form (\ref{121}), ${\cal L}_{TOT}$
is manifestly gauge invariant.

After solving the equations of motion for the soldering fields we can write, 
\begin{equation}  \label{133}
{\cal L}_g ={\frac{1}{2}} \sqrt{-g}g^{\alpha\beta}
\partial_\alpha\Phi\partial_\beta\Phi\;\;.
\end{equation}

\noindent where, in the above expression we have introduced the metric tensor
density 
\begin{eqnarray}  \label{134}
\sqrt{-g}g^{--}&=&-4{\frac{\lambda_{++}}{\Delta}}\,,  \nonumber \\
\sqrt{-g}g^{++}&=&-4{\frac{\lambda_{--}}{\Delta}}\,,  \nonumber \\
\sqrt{-g}g^{+-}&=&-{\frac{2}{\Delta}}(1+\lambda_{++}\lambda_{--}) \,,
\end{eqnarray}

\noindent where $\Delta=2(\lambda_{++}\lambda_{--}-1)$ and 
\begin{equation}  \label{135}
\Phi={\frac{1 }{\sqrt{2}}}(\rho-\varphi)\,.
\end{equation}

We observe that in two dimensions $\sqrt{-g}g^{\alpha\beta}$ needs only two
parameters to be defined in a proper way. As it should be, $det(\sqrt{-g}%
g^{\alpha\beta})=-1$. We also note that, because of conformal invariance, we
cannot determine $g_{\alpha\beta}$ itself. We could, therefore, think of $%
{\cal L}_{TOT}$ as an effective theory, which represents a scalar boson $\Phi
$ in a gravitational background. It can be shown \cite{adw} that the action (%
\ref{133}) can be made invariant under the full group of diffeomorphism.
Hence, we can easily see that, in terms of symmetry, the new theory is
bigger than the old one. This new theory can be interpreted as a
constructive interference of symmetries. However, solving the equations of
motion for the multipliers, we can see that, in fact, this field has no
dynamics. This characterizes a nonmover field, a noton, introduced by Hull 
\cite{hull} to cancel out the gravitational anomaly of the Siegel model.

\section{The master action}

In this section we will propose a master action which represents, as a function of arbitrary parameters, several theories for the Siegel gauged model. In the second
part we have accomplished the soldering of opposite chiral versions of this
master action and applied the final result, i.e., the 
soldered action, on several models for the self-dual theory to make an interference analysis
of the covariance of the new theories.

\subsection{The generalized gauged Siegel model}

Let us now construct a class of generalized actions for Abelian chiral
bosons coupled to a gauge field for each chirality, i.e., for the coupled
leftons (${\cal L}_L$) and rightons (${\cal L}_R$). We will call it the
generalized gauged Siegel model (GGSM), 
\begin{mathletters}
\begin{eqnarray}
{\cal L}^{(0)}_{L}\,&=&\,(\,\partial_+\,\phi\,+\,a_1\,A_+\,)\,(\,\partial_-\,%
\phi\,+\,a_2\,A_-\,)\, \nonumber \\
&+&\, \lambda_{++}\,(\,\partial_-\,\phi\,+\,a_3\,A_-\,)^2
\label{eqa}
\end{eqnarray}
\begin{eqnarray}
{\cal L}^{(0)}_{R}\,&=&\,(\,\partial_+\,\rho\,+\,b_1\,A_+\,)\,(\,\partial_-\,%
\rho\,+\,b_2\,A_-\,)\, \nonumber \\
&+&\,
\lambda_{--}\,(\,\partial_+\,\rho\,+\,b_3\,A_+\,)^2 \;\;,  \label{eqb}
\end{eqnarray}
\end{mathletters}
where $a_i,\,b_i\,(i=1,2,3)$ are parameters that define the theory studied
and $A_\pm$ are the vector field components. We will see below that making simple
substitutions of these parameters we can obtain several gauged forms of the
Siegel theory that appear in the literature.  It is important to observe the 
difference between the vector fields $A_\pm$ above and the soldering fields $B_\pm$ of 
equations (\ref{121}).  The $A$-fields are external (or background) fields and hence one does not consider the variation (and extrema) of the actions under the variations of these fields.
The last are the auxiliary fields, as mentioned above, which helps 
in the soldering process and will be naturally eliminated by solving its equations of motion.

Following the steps of the soldering formalism studied in the last section,
we can start considering the variation of the Lagrangians under the usual
transformations, 
\begin{equation}  \label{31}
\delta\,\phi\,=\,\delta\,\rho\,=\,\alpha \;\; \mbox{and} \;\;
\delta\,A_{\mu}\,=\,\partial_{\mu}\,\alpha
\end{equation}
where $\mu=+,-$. Now, consider that this symmetry is a global one with, obviously, a global parameter $\alpha$ so that the above transformations take the form
\begin{equation}  \label{32}
\delta\,\phi\,=\,\delta\,\rho\,=\,\alpha \;\; \mbox{and} \;\;
\delta\,A_{\mu}\,=\,0\;\;.
\end{equation}
Remember that the soldering process consists in lifting the gauging of a global symmetry to its local version.  Hence we will consider from now on the transformations (\ref{32}) as local.
Let us continue with the procedure writing only the main steps of the procedure.

In terms of the Noether currents we can construct 
\begin{equation}
\delta {\cal L}^{(0)}_{L,R}\,=\,J^{\mu}_{\phi,\rho}\,\partial_{\mu}\,\alpha%
\;\;,
\end{equation}
where 
\begin{eqnarray}
J^{+}_{\phi}&=&a_2\,A_-  \nonumber \\
J^{-}_{\phi}&=&2\,\partial_+\,\phi\,+\,a_1\,A_+\,+\,2\,\lambda_{++}\,(\,%
\partial_-\,\phi\,+\,a_3\,A_-\,)  \nonumber \\
J^{+}_{\rho}&=&b_2\,A_-  \nonumber \\
J^{-}_{\rho}&=&2\,\partial_-\,\rho\,+\,b_2\,A_-\,+\,2\,\lambda_{--}\,(\,%
\partial_+\,\rho\,+\,b_3\,A_+\,) \;\;.
\end{eqnarray}

\noindent The next iteration, as seen above, can be performed introducing
auxiliary fields, the so-called soldering fields 
\begin{equation}
{\cal L}^{(1)}_{L,R}\,=\,{\cal L}^{(0)}_{L,R}\,-\,B_{\mu}\,J^{\mu}_{\phi,%
\rho}\;\;,
\end{equation}

\noindent and one can easily see that the gauge variation of the GGSM is

\begin{equation}  \label{34}
\delta {\cal L}^{(1)}_{L,R}\,=\,-\,B_{\mp}\,\delta\,B_{\pm}\,-\,\lambda_{\pm%
\pm}\,\delta\, B^2_{\mp}\;\;.
\end{equation}
Let us define the variation of $B_{\pm}$ as 
\begin{equation}
\delta B_{\pm}=\partial_{\pm}\alpha\;\;,
\end{equation}
and we see that the variation of ${\cal L}^{(1)}_{L,R}$ does not depend
neither on $\phi$ nor $\rho$. Hence, as explained in the last section, we
can construct the final (soldered) Lagrangian as 
\begin{eqnarray}  \label{35}
{\cal L}&=&\,{\cal L}_{L}\,\oplus\,{\cal L}_{R}  \nonumber \\
&=&{\cal L}^{(1)}_{L}\,+\,{\cal L}^{(1)}_{R}\,+\,2\,B_+\,B_-\,+
\,\lambda_{++}\,B^2_-\, +\,\lambda_{--}\,B^2_+  \nonumber \\
&=&\,(\,\partial_+\,\phi\,+\,a_1\,A_+\,)\,(\,\partial_-\,\phi\,+\,a_2\,A_-%
\,)\, \nonumber \\
&+&\, \lambda_{++}\,(\,\partial_-\,\phi\,+\,a_3\,A_-\,)^2  \\
&+&
(\,\partial_+\,\rho\,+\,b_1\,A_+\,)\,(\,\partial_-\,\rho\,+\,b_2\,A_-\,)\, \nonumber \\
&+&
\, \lambda_{--}\,(\,\partial_+\,\rho\,+\,b_3\,A_+\,)^2  \nonumber \\
&-&B_{\mu}\,J^{\mu}_{\phi}\,-\,
B_{\mu}\,J^{\mu}_{\rho}\,+\,2\,B_+\,B_-\,+\,\lambda_{++}\,B^2_-\, \nonumber \\
&+&\,\lambda_{--}\,B^2_+ \;\;, \nonumber                  
\end{eqnarray}
which remains invariant under the combined transformations (\ref{31}) and (%
\ref{34}). Following the steps of the algorithm depicted in the last
section, we have to eliminate the soldering fields solving their equations
of motion which results in 
\begin{equation}
B_{\pm}\,=\,\frac{J^{\mp}\,-\,\lambda_{\pm\pm}\,J^{\pm}}{2\,(1\,-\,\lambda)}%
\;\;,
\end{equation}
where $\lambda=\lambda_{++}\,\lambda_{--}$ and $J^{\pm}=J^{\pm}_{\phi}+J^{%
\pm}_{\rho}$.

Substituting it back in (\ref{35}) we have the final soldered action
\begin{eqnarray}  \label{37}
{\cal L}\,&=&\,{\frac{1 }{2}}\,\sqrt{-g}\,g^{\mu\nu}\,\partial_{\mu}\,\Phi\,%
\partial_{\nu}\,\Phi  \nonumber \\
&+&\,{\frac{1}{{1-\lambda}}}\left\{\,
(a_1\,+\,b_1\,\lambda\,-\,2\,\lambda\,b_3)\,\partial_-\,\Phi\,A_+\, \right. \nonumber \\
&+& \left. (\,2\,\lambda\,a_3\,-\,a_2\,\lambda\,-\,b_2)\,\partial_+\,\Phi\,A_-\, \right.
\nonumber \\
&+& \left.
\,\lambda_{++}\,(2\,a_3\,-\,a_2\,-\,b_2)\,\partial_-\,\Phi\,A_-\, \right. \nonumber \\
&+& \left. \lambda_{--}(a_1\,+\,b_1\,-\,2\,b_3)\,\partial_+\,\Phi\,A_+ \right. 
\nonumber \\
&+& \left.
\,C_1\,\lambda_{++}\,A^2_-\,+\,C_2\,\lambda_{--}\,A^2_+\,+\,C_{\lambda}%
\,A_+\,A_- \,\right\}
\end{eqnarray}
where the new compound field are defined as $\Phi=\phi\,-\,\rho$. The new
parameters are 
\begin{eqnarray}
C_1&=&a_3^2\,-\,b_2\,a_3\,+\,{\frac{1 }{4}}(a_2\,+\,b_2)^2  \nonumber \\
C_2&=&b_3^2\,-\,b_3\,a_1\,+\,{\frac{1 }{4}}(a_1\,+\,b_1)^2  \nonumber \\
C_{\lambda}&=&({\frac{1 }{2}}\,-\,\lambda)\,a_1\,a_2\,+\,({\frac{1 }{2}}%
\,-\,\lambda)\,b_1\,b_2\,-\, {\frac{1 }{2}}\,(a_1\,b_2\,+\,b_1\,a_2) 
\nonumber \\
&+&\,[(a_2\,+\,b_2)\,b_3\,+\,(a_1\,+\,b_1)\,a_3\,-\,
2\,a_3\,b_3]\,\lambda\, \nonumber \\
&-&\,a_2\,a_3\,\lambda_{++}\,-\,b_1\,b_3\,\lambda_{--}%
\;\;.
\end{eqnarray}
and the metric is 
\begin{equation}  \label{metrica}
{\frac{1 }{2}}\,\sqrt{-g}\,g^{\mu\nu} = {\frac{1 }{{2\,(1-\lambda)}}}%
\,\left( 
\begin{array}{cc}
2\lambda_{--} & 1+\lambda \\ 
1+\lambda & 2\lambda_{++}
\end{array}
\right)
\end{equation}
which reminds the gravitational feature of the soldered action of the two
Siegel modes. We can note that the action (\ref{37}) is covariant. Hence, in
this case, we have that the covariance of the generalized gauged Siegel
action is maintained.  This general action form will allow us to apply it to the
various gauged theories for the chiral boson with second order constraint.
This will be accomplished next.

\subsection{The self-dual models}

In this section we will analyze five kinds of theories in the light of the
soldering formalism. The first of them is the well known Siegel's action 
\cite{siegel}, studied in section II. It has been used to demonstrate the
validity of the general soldered action (\ref{37}). The second example, which is not a new result also, will be a coupling of the chiral boson with a gauge
field. We are talking, in this case, about the Gates and Siegel gauged
action \cite{gs}. The new results will appear with the next three models. 
We will
use three models well known in the literature: the one derivative gauged
model, the massless Bellucci, Golterman and Petcher model and the Frishman
and Sonnenschein model.

\subsubsection{Siegel's model}

It is easy to see that to obtain the expression (\ref{03}) we have to fix
the parameters with the following values: 
\begin{equation}
a_{i}\,=\,b_{i}\,=\,0
\end{equation}
where $i=1,2,3$. Hence, substituting these values in the expression (\ref{37}%
) it follows that 
\begin{eqnarray}
{\cal L}_{TOT}\, &=&\,{\frac{1}{{1-\lambda }}}\left\{ \left( 1+\lambda
_{++}\lambda _{--}\right) \partial _{-}\Phi \partial _{+}\Phi \,+\,\lambda
_{++}\left( \partial _{-}\Phi \right) ^{2} \right. \nonumber \\
&+& \left. \lambda _{--}\left( \partial
_{+}\Phi \right) ^{2}\right\}   \nonumber \\
&=&{\frac{1}{2}}\,\sqrt{-g}\,g^{\mu \nu }\,\partial _{\mu }\,\Phi \,\partial
_{\nu }\,\Phi 
\end{eqnarray}
where ${\frac{1}{2}}\,\sqrt{-g}\,g^{\mu \nu }$, from now on, is written like in (%
\ref{metrica}). This action represents, naively, a scalar field
immersed in a gravitational background. However, as we have stressed in
section II, this expression also represents the noton action.

\subsubsection{Gates and Siegel's model}

Gates and Siegel \cite{gs} have studied the interactions of leftons and
rightons with external vector fields including the supersymmetric and the
non-Abelian cases. The soldering of this model has been obtained already
in \cite{aw}, but as a further test for our GGSM, let us write 
\begin{eqnarray}
{\cal L}_{GS}^{\phi}\,&=&\,(\,\partial_-\,\phi\,+\,2\,A_-\,)\,(\,\partial_+\,%
\phi\,)\,+\, \lambda_{++}\,(\,\partial_-\,\phi\,+\,A_-\,)^2  \nonumber \\
{\cal L}_{GS}^{\rho}\,&=&\,(\,\partial_+\,\rho\,+\,2\,A_+\,)\,(\,\partial_-\,%
\rho\,)\,+\, \lambda_{--}\,(\,\partial_+\,\rho\,+\,A_+\,)^2 \nonumber \\
& &                  
\end{eqnarray}
and the correspondence with (\ref{eqa}) is direct 
\begin{equation}
a_2\,=\,b_1\,=\,2 \;\; ; \;\; a_1\,=\,b_2\,=\,0 \;\;;\;\;
a_3\,=\,b_3\,=\,1\;.
\end{equation}

The soldered action is, using (\ref{37}), 
\begin{equation}  \label{II11}
{\cal L}_{TOT}\,=\,{\frac{1 }{2}}\,\sqrt{-g}\,g^{\mu\nu}\,\partial_{\mu}\,%
\Phi\,\partial_{\nu}\,\Phi \,-\,2\,A_{-}A_{+}\;\;,
\end{equation}
confirming the result in \cite{aw}. We can note that the covariance has not
been broken.

The physical meaning of (\ref{II11}) can be appreciated by eliminating the
multipliers and using the symmetry induced by the soldering \cite{wotzasek},
showing that it represents the action for the noton. In fact (\ref{II11}) is
basically the action proposed by Hull \cite{hull} as a candidate for
canceling the Siegel anomaly. This field carries a representation of the
full diffeomorphism group \cite{hull} while its chiral (Siegel) component
carry the representation of the chiral diffeomorphism. Observe the complete
disappearance of the dynamical sector due to the destructive interference
between the leftons and the rightons. This happens because we have
introduced only one soldering field to deal with both the dynamics and the
symmetry. To recover dynamics we need to separate these sectors and solder
them independently, as stressed in \cite{aw}.

\subsubsection{One derivative gauged model}

This gauged form was introduced in \cite{dd1}, where only one kind of
derivative were gauged, 
\begin{eqnarray}
\label{47}
{\cal L}_{OD}^{\phi }\, &=&\,(\,\partial _{+}\,\phi
\,+\,2\,e\,A_{+}\,)\,(\,\partial _{-}\,\phi \,)\,+\,\lambda
_{++}\,(\,\partial _{-}\,\phi \,)^{2}  \nonumber \\
{\cal L}_{OD}^{\rho }\, &=&\,(\,\partial _{-}\,\rho
\,+\,2\,e\,A_{-}\,)\,(\,\partial _{+}\,\rho \,)\,+\,\lambda
_{--}\,(\,\partial _{+}\,\rho \,)^{2}
\end{eqnarray}
Hence, immediately we have the correspondence with the equations (\ref{eqa})
and (\ref{eqb}) through the choice
\begin{equation}
a_{2}\,=\,a_{3}\,=\,0\;\; ;\;\; b_{1}\,=\,b_{3}\,=\,0\;\;;\;\;
a_{1}\,=\,b_{2}\,=\,2\,e\;\;.
\end{equation}
and 
\begin{eqnarray}
& &{\cal L}_{TOT}\, =\,{\frac{1}{2}}\,\sqrt{-g}\,g^{\mu \nu }\,\partial _{\mu
}\,\Phi \,\partial _{\nu }\,\Phi \, \nonumber \\
& &+\,{\frac{1}{{1-\lambda }}}\,\left[
\,-\,2\,e\,(\partial _{-}\,\Phi \,A_{+}\,-\,\partial _{+}\,\Phi
\,A_{-})\,\right.   \nonumber \\
&&\left. -\,2\,e\,(\lambda _{++}\,\partial _{-}\,\Phi \,A_{-}\,-\,\lambda
_{--}\,\partial _{+}\,\Phi \,A_{+}\,)\, \right. \nonumber \\
&&+ \left. \,e^{2}\,(\,\lambda
_{++}\,A_{-}^{2}\,+\,\lambda _{--}\,A_{+}^{2}\,-\,2\,A_{+}\,A_{-}\,)\right]
\;\;.
\end{eqnarray}
In this case we can note that the decoupling of the vector fields has not occurred.

The final action is explicitly covariant, showing that the soldering procedure 
did not provide the break of covariance.  We can classify this case as constructive
interference of covariances, since equations (\ref{47}) are covariant also.

\subsubsection{The gauged massless Bellucci, Golterman and Petcher model}

The form of this gauged chiral boson action is 
\begin{eqnarray}
{\cal L}_{BGP}^{\phi}\,=\,(\,\partial_+\,\phi\,)\,(\,\partial_-\,\phi\,&+&
\,e\,A_-\,)\, \nonumber \\
&+&\, \lambda_{++}\,(\,\partial_-\,\phi\,+\,e\,A_-\,)^2 
\nonumber \\
{\cal L}_{BGP}^{\rho}\,=\,(\,\partial_-\,\rho\,)\,(\,\partial_+\,\rho\,&+&
\,e\,A_+\,)\, \nonumber \\
&+&\, \lambda_{--}\,(\,\partial_+\,\rho\,+\,e\,A_+\,)^2                            
\end{eqnarray}
hence 
\begin{equation}
a_1\,=\,b_2\,=\,0 \;\; ; \;\; a_2\,=\,a_3\,=\,b_1\,=\,b_3\,=\,e\;\;.
\end{equation}
and the final action reads, 
\begin{eqnarray}
& &{\cal L}_{TOT}\,=\,{\frac{1 }{2}}\,\sqrt{-g}\,g^{\mu\nu}\,\partial_{\mu}\,%
\Phi\,\partial_{\nu}\,\Phi\, \nonumber \\
&+&\, {\frac{1 }{{1-\lambda}}}\,\left[\,e\,\lambda%
\,(\partial_+\,\Phi\,A_-\,-\,\partial_-\,\Phi\,A_+\,) \right.  \nonumber \\
&+&\left.
e\,\lambda_{++}\,\partial_-\,\Phi\,A_-\,-\,e\,\lambda_{--}\,(1\,+\,\lambda)\,%
\partial_+\,\Phi\,A_+\, \right. \nonumber \\
&+& \left. \, {\frac{5}{4}}\,e^2\,(\,\lambda_{++}\,A_-^2\,+\,%
\lambda_{--}\,A_+^2\,) \right.  \nonumber \\
&-& \left. e^2\, ({\frac{1}{2}}\,-\,\lambda_{++}\,-\,\lambda_{--}\,)\,A_+%
\,A_-\,\right]\;\;,
\end{eqnarray}
it is easy to see that last two terms break the covariance.  Hence, in this case
we have clearly a destructive interference of covariances.

\subsubsection{The Frishman and Sonnenschein model}

The chiral actions developed in \cite{fs} are 
\begin{eqnarray}
{\cal L}_{FS}^{\phi}\,&=&\,(\,\partial_+\,\phi\,)\,(\,\partial_-\,\phi\,)\,+%
\,
\lambda_{++}\,(\,\partial_-\,\phi\,)^2\,+\,\partial_+\,\phi\,A_-\, \nonumber \\
&-&\,\partial_-\,\phi\,A_+,  \nonumber \\
{\cal L}_{FS}^{\rho}\,&=&\,(\,\partial_-\,\rho\,)\,(\,\partial_+\,\rho\,)\,+%
\,
\lambda_{--}\,(\,\partial_+\,\rho\,)^2\,+\,\partial_-\,\rho\,A_+\, \nonumber \\
&-&\,\partial_+\,\rho\,A_-\;\;,
\end{eqnarray}
and identifying the parameters, 
\begin{equation}
a_1\,=\,b_2\,=\,-\,1 \;\; ; \;\; a_2\,=\,b_1\,=\,1 \;\; ;\;\;
a_3\,=\,b_3\,=\,0\;\;,
\end{equation}
we can construct the soldered action as 
\begin{eqnarray}
{\cal L}_{TOT}\,&=&\,{\frac{1 }{2}}\,\sqrt{-g}\,g^{\mu\nu}\,\partial_{\mu}\,%
\Phi\,\partial_{\nu}\,\Phi\,+\,
\epsilon^{\mu\nu}\,\partial_{\mu}\Phi\,A_{\nu}\, \nonumber \\
&+&\,{\frac{{1-2\lambda} }{{%
1-\lambda}}}\,A_+\,A_-\;\;,
\end{eqnarray}
where $\epsilon^{+-}=1$.  

Now we have a constructive interference of covariance, since, the soldered action
is explicitly covariant.

\section{The Lorentz invariance analysis}

Let us now fix conditions over the parameters in order to respect a Lorentz
invariance. In other words we mean that we will fix conditions such
that the constraints valid in one inertial reference system are valid in the another
one. To make this, we will perform the Lorentz rotation \cite{dd1}.   This will
be done in the corresponding FJ version of the GGSM  proposed above.

The generalized gauged Siegel model as we already know, is 
\begin{eqnarray}
{\cal L}_{GGSM}\,&=&\,(\,\partial_+\,\phi\,+\,k_1\,A_+\,)\,(\,\partial_-\,\phi%
\,+\,k_2\,A_-\,)\, \nonumber \\
&+&\,\lambda_{++}\,(\,\partial_-\,\phi\,+\,k_3\,A_-\,)^2\;\;.
\end{eqnarray}

\noindent The canonical momentum conjugated to $\phi$ is 
\begin{eqnarray}  \label{vinculo}
\pi_{\phi}\,&=&\,{\phi^{\prime}}\,+\,{\frac{1 }{2}} k_1\,(\,A_0\,+\,A_1\,)%
\,+\,{\frac{1 }{2}}\,(\,k_2\,-\,k_3\,)\, (\,A_0\,-\,A_1\,)  \nonumber \\
&=&{\phi^{\prime}}\,+\,{\frac{1 }{2}} \,(\,k_1\,+\,k_2\,-\,2\,k_3\,)\,A_0  \nonumber \\
&+&\,{\frac{1 }{2}}\, (\,k_1\,-\,k_2\,+\,2\,k_3\,)\,A_1\;\;,
\end{eqnarray}
and this is the generalized chiral constraint.

Using the first-order formalism of Faddeev and Jackiw with this momentum, we
can construct a first-order Lagrangian density, 
\begin{eqnarray}
& &{\cal L}_{GGSM}\,=\,\dot{\phi}\,\phi^{\prime}\,-\,{\phi^{\prime}}^2\,+\,%
\frac{A_0^2}{2}\, (\,k_1\,k_2\,-\,k_3^2\,) \, \nonumber \\
&&-\,\frac{A_1^2}{2}\,(\,k_1\,k_2\,+\,k_3^2\,)   \\
& &+\, \frac{A_0^2}{2}\,[\,(\,k_1\,+\,k_2\,-\,2\,k_3\,)\,\dot{\phi}\,-\,
(\,k_1\,-\,k_2\,-\,2\,k_3\,)\,\phi^{\prime}\,]  \nonumber \\
& &+\,\frac{A_1^2}{2}\,[\,(\,k_1\,-\,k_2\,+\,2\,k_3\,)\,\dot{\phi}
-\,(\,k_1\,+\,k_2\,+\,2\,k_3\,)\,\phi^{\prime}\,] \nonumber 
\end{eqnarray}

\noindent which is a constrained one. To verify Lorentz invariance we have
to note if the constraints are preserved from one inertial reference system to the
other. To do this we have to apply the Lorentz rotation on the generalized
chiral constraint. Constructing the rotation matrices as, 
\begin{equation}  \label{uma}
\left( 
\begin{array}{c}
\pi \\ 
\phi^{\prime}
\end{array}
\right) \rightarrow \,\left( 
\begin{array}{cc}
cosh\,\varphi & sinh\,\varphi \\ 
sinh\,\varphi & cosh\,\varphi
\end{array}
\right) \,\left( 
\begin{array}{c}
\tilde{\pi} \\ 
\tilde{\phi^{\prime}}
\end{array}
\right)
\end{equation}
and 
\begin{equation}  \label{duas}
\left( 
\begin{array}{c}
A_0 \\ 
A_1
\end{array}
\right) \rightarrow \,\left( 
\begin{array}{cc}
cosh\,\varphi & sinh\,\varphi \\ 
sinh\,\varphi & cosh\,\varphi
\end{array}
\right) \,\left( 
\begin{array}{c}
\tilde{A_0} \\ 
\tilde{A_1}
\end{array}
\right) \;\;,
\end{equation}
and we have relations between the old fields and the new (tilde) fields.
Writing the eq. (\ref{vinculo}) in a convenient way 
\begin{equation}  \label{vinculo2}
\pi_\phi\,=\,\phi^{\prime}\,+\,{\frac{C_1 }{2}}\,A_0\,+\,{\frac{C_2 }{2}}%
\,A_1\;\;,
\end{equation}
where $C_1=k_1\,+\,k_2\,-\,2\,k_3$ and $C_2=k_1\,-\,k_2\,+\,2\,k_3$.

After a little algebra, where we have provided the substitution of (\ref{uma}%
) and (\ref{duas}) in (\ref{vinculo2}), we can write, 
\begin{eqnarray}
C_1\,cosh\,\varphi\,+\,C_2\,sinh\,\varphi&=&C_1\,(cosh\,\varphi\,-\,sinh\,%
\varphi)  \nonumber \\
C_1\,sinh\,\varphi\,+\,C_2\,cosh\,\varphi&=&C_2\,(cosh\,\varphi\,-\,sinh\,%
\varphi)\;\;.
\end{eqnarray}
Solving this system we can say that the generalized chiral constraints are
Lorentz invariant, if 
\begin{equation}  \label{15}
C_1\,=\,-\,C_2\;\;.
\end{equation}
In other words we can say that with this solution the constraint is
independent of the reference system.

Solving equation (\ref{15}) we have that 
\begin{equation}
k_1\,=\,0\;\;.
\end{equation}
With this result, we conclude that we can only gauge terms with the same
light-cone variables, i.e., 
\begin{equation}
{\cal L}\,=\,(\,\partial_+\,\phi\,)\,(\,\partial_-\,\phi\,+\,k_2\,A_-\,)\,+%
\, \lambda_{++}\,(\,\partial_-\,\phi\,+\,k_3\,A_-\,)^2 \;\;,
\end{equation}
which corroborates the results for the gauging of the FJ model.

At this point it is interesting to remark that, in the original proposal of 
this method for verifying the relativistic invariance using the Lorentz rotation \cite{dd1,dd2},  it was supposed
that the invariance should be imposed, and this had lead to some criticisms \cite{clo}.
Now we can see that in this approach, in fact, there is no need of {\it ad hoc} impositions.

%\section{The soldering formalism, covariance and Lorentz invariance}

\section{The chiral Schwinger model with generalized constraint}

In reference \cite{dd1} the Lorentz rotation technique was used in the
bosonized form of the  chiral Schwinger model with a generalized constraint 
\begin{equation}
\Omega\,=\,\pi_{\phi}\,-\,\alpha\,\phi^{\prime}
\end{equation}
imposed on the first-order Lagrangian to determinate conditions on $\alpha$
such that we have a Lorentz invariant final theory.

Now we will disclose the conditions on $\alpha$ in the soldered action in
order to have a covariant model. To begin with, let us write both chiralities of the effective Lagrangian 
\cite{dd1}, 
\begin{mathletters}
\begin{eqnarray}
{\cal L}_{\alpha}^{\phi}\,&=&\,-\,\alpha\,\dot{\phi}\,\phi^{\prime}\,-\,{\frac{%
1}{2}}\,(\alpha^2\,+\,1)\, {\phi^{\prime}}^2\, \nonumber \\
&+&\,(\alpha\,+\,1)\,e\,\phi^{%
\prime}\,(\,A_0\,-\,A_1\,)\,-\,{\frac{1}{2}}\,e^2\, (\,A_0\,-\,A_1\,)\, \nonumber \\
&+&\,{\frac{a}{2}}\,A_{\mu}^2  \label{eq2a}
\end{eqnarray}
\begin{eqnarray}
{\cal L}_{\alpha}^{\rho}\,&=&\,\alpha\,\dot{\rho}\,\rho^{\prime}\,-\,{\frac{1}{%
2}}\,(\alpha^2\,+\,1)\, {\rho^{\prime}}^2\, \nonumber \\
&+&\,(\alpha\,+\,1)\,e\,\rho^{%
\prime}\,(\,A_0\,+\,A_1\,)\,-\,{\frac{1}{2}}\,e^2\, (\,A_0\,+\,A_1\,)\, \nonumber \\
&+&\,{\frac{b}{2}}\,A_{\mu}^2  \label{eq2b}
\end{eqnarray}
where, in \cite{dd1}, to produce a Lorentz covariant theory, $\alpha$ is the
solution of the equation, 
\end{mathletters}
\begin{equation}
(\alpha^2\,+\,1)\,\phi^{\prime}\,-\,(\,g_1\,\alpha\,+\,g_2\,)\,A_0\,-\,
(\,g_2\,\alpha\,+\,g_1\,)\,A_1\,=\,0\;\;.
\end{equation}
The parameters $g_1$ and $g_2$ are $g_1=g_2=e\;\;(g_1=-g_2=e)$ for right
(left)-handed chiral Schwinger model.

Now, to perform an interference analysis, we have to impose the gauge
transformations 
\begin{equation}
\left( 
\begin{array}{c}
\phi \\ 
\rho
\end{array}
\right) \;\; \rightarrow \;\; \left( 
\begin{array}{c}
\phi \\ 
\rho
\end{array}
\right) \:\:\:+\:\:\:\xi \:\left( 
\begin{array}{c}
1 \\ 
1
\end{array}
\right) 
\end{equation}
which, following the soldering mechanism, has as Noether currents, 
\begin{eqnarray}  \label{933}
J_\phi^{0}&=&J_\rho^{0}\,=\,0  \nonumber \\
J_\phi^{1}&=&-\,2\,\alpha\,\dot{\phi}\,-\,(\alpha^2\,+\,1)\,\,\phi^{\prime}%
\,+\, e\,(\alpha\,+\,1)\,(\,A_0\,-\,A_1\,)  \nonumber \\
J_\rho^{1}&=&2\,\alpha\,\dot{\rho}\,-\,(\alpha^2\,+\,1)\,\,\rho^{\prime}\,+%
\, e\,(\alpha\,-\,1)\,(\,A_0\,+\,A_1\,)
\end{eqnarray}

Introducing the soldering fields and eliminating them by solving their
equations of motion and substituting back into the contact terms of the
action, we have a final soldered action 
\begin{eqnarray}  \label{final}
& &{\cal L}_{FINAL}=\,-\,{\frac{1}{4}}\,(\alpha^2\,+\,1)\,{\Phi^{\prime}}^2\,+\,%
\frac{\alpha^2}{\alpha^2\,+\,1} \dot{\Phi}^2 \, \nonumber \\
&+&\,\frac{2\,\alpha^2\,e}{
\alpha^2\,+\,1}\,\dot{\Phi}\,A_0\,+\,e\,\alpha\,\Phi^{\prime}\,A_1 \,-\,%
\frac{2\,\alpha\,e}{\alpha^2\,+\,1}\,\dot{\Phi}\,A_1  \nonumber \\
&+&\,e\,\Phi^{\prime}\,A_0\,+\,e^2\,\left[\,\frac{\alpha^2}{\alpha^2\,+\,1%
} \,+\,{\frac{1}{2}}\,(a\,+\,b)\,-\,1\,\right]\,A_0^2  \nonumber \\
&+&\,e^2\,\left[\,\frac{1}{\alpha^2\,+\,1}\,+\,{\frac{1}{2}}%
\,(a\,+\,b)\,-\,1\,\right]\,A_1^2 \, \nonumber \\
&-&\,\frac{2\,\alpha\,e^2}{\alpha^2\,+\,1}%
\,A_0\,A_1\;\;,
\end{eqnarray}
remembering that $\Phi=\phi\,-\,\rho$, as usual. We can easily see that $%
{\cal L}_{FINAL}$ does not describe a constrained system. The soldering
procedure has broken the constraint feature of the system. This fact is
contrary to the feature of eqs. (\ref{eq2a}) and (\ref{eq2b}), which are
constrained Lagrangians. Hence, we
will ask which conditions $\alpha$ must obey in order to preserve the
manifest covariance and consequently the Lorentz invariance.

\subsection{Manifest covariance}

To obtain the manifest covariance, it is easy to see that, in (\ref{final}) $%
\alpha$ have to satisfy the following set of equations, 
\begin{mathletters}
\begin{equation}
{\frac{1}{4}}\,(\alpha^2\,+\,1)=\frac{\alpha^2}{\alpha^2\,+\,1}  \label{eq3a}
\end{equation}
\begin{equation}
\frac{2\,\alpha^2\,e}{\alpha^2\,+\,1}=-\,\alpha  \label{eq3b}
\end{equation}
\begin{equation}
\frac{2\,\alpha}{\alpha^2\,+\,1}=-\,1  \label{eq3c}
\end{equation}
\begin{equation}
-\,\frac{\alpha^2}{\alpha^2\,+\,1}\,=\frac{1}{\alpha^2\,+\,1}\,
\label{eq3d}
\end{equation}
\begin{equation}
\frac{\alpha}{\alpha^2\,+\,1}=0\;\;.  \label{eq3e}
\end{equation}

Analyzing the solution $\alpha =0$ of equation (\ref{eq3e}) we can easily see that it is not
compatible with equations (\ref{eq3a}), (\ref{eq3c}) and (\ref{eq3d}). We
can observe, for instance, that equation (\ref{eq3d}) presents a complex solution also, $%
\alpha =\pm i$. Hence, our soldered action is not manifestly covariant at
all. It is interesting to notice that the massive terms for the gauge fields
of the actions (\ref{eq2a}) and (\ref{eq2b}) have not influenced the final
result. The condition to impose covariance on the gauge fields terms, i.e., $%
\alpha =0$, at the same time breaks the covariance of the action
independently of the gauge field massive terms. This result supplements the
one found in \cite{abw}, where the soldering of two massless chiral
Schwinger models generates a massive particle.

\subsection{Lorentz invariance}

As we saw in the last section, the action (\ref{final}) is not constrained.
So, to impose conditions on $\alpha$ to verify if our final (soldered)
action, eq. (\ref{final}), is Lorentz invariant through a Lorentz rotation,
we have to make a direct comparison term by term. The first step is to
rewrite the action (\ref{final}) as 
\end{mathletters}
\begin{eqnarray}
{\cal L}\,&=&\,a_1\,{\phi^{\prime}}^2\,+\,a_2\,\dot{\phi}^2\,+\,a_3\,\dot{\phi}%
\,A_0\,+\, e\,\alpha\,{\phi^{\prime}}\,A_1\,+\,e\,{\phi^{\prime}}%
\,A_0\, \nonumber \\
&+&\,a_5\,A_0^2\,+\,a_6\,A_1^2\,+\,a_7\,A_0\,A_1
\end{eqnarray}
where 
\begin{eqnarray}  \label{coeficientes}
a_1&=&-\,{\frac{1}{4}}\,(\,\alpha^2\,+\,1\,)  \nonumber \\
a_2&=&\frac{\alpha^2}{\alpha^2+1}  \nonumber \\
a_3&=&-\,\frac{2\,\alpha^2}{\alpha^2+1}  \nonumber \\
a_4&=&-\,\frac{2\,e\,\alpha^2}{\alpha^2+1}   \\
a_5&=&e^2\,\left[\frac{\alpha^2}{\alpha^2+1}\,+\,{\frac{1}{2}}%
\,(\,a\,+\,b\,)\,-\,1\,\right]  \nonumber \\
a_6&=&e^2\,\left[\frac{1}{\alpha^2+1}\,+\,{\frac{1}{2}}\,(\,a\,+\,b\,)\,-\,1%
\,\right]  \nonumber \\
a_7&=&-\,\frac{2\,e^2\,\alpha^2}{\alpha^2+1} \nonumber
\end{eqnarray}

Following the Lorentz rotation procedure, we have to establish the matrix
relations between the old and the new (tilde) fields through the
construction of Lorentz rotation matrix, 
\begin{equation}
\left( 
\begin{array}{c}
\dot{\Phi} \\ 
\Phi^{\prime}
\end{array}
\right) \;\; \rightarrow \;\; \left( 
\begin{array}{cc}
cosh\,\theta & sinh\,\theta \\ 
sinh\,\theta & cosh\,\theta
\end{array}
\right)\;\; \left( 
\begin{array}{c}
\tilde{\dot{\Phi}} \\ 
\tilde{\Phi}^{\prime}
\end{array}
\right)\;\;.
\end{equation}
and for the gauge fields components, (\ref{duas}).

\bigskip \bigskip

Finally, after these substitutions, the transformed Lagrangian is 
\begin{eqnarray}
& &{\cal L}_{FINAL}\,=\,(\,a_2\,x^2\,+\,a_1\,y^2\,)\,\tilde{\dot{\phi}}%
^2\,+\, (\,a_2\,y^2\,+\,a_1\,x^2\,)\,\tilde{\phi^{\prime}}^2\, \nonumber \\
&+&\,
2\,(\,a_1\,+\,a_2\,)\,x\,y\,\tilde{\dot{\phi}}\,\tilde{\phi^{\prime}} 
\nonumber \\
&+&\,(\,a_3\,y^2\,+\,e\,y^2\,+\,e\,\alpha\,x\,y\,+\,a_4\,x^2\,)\,\tilde{%
\dot{\phi}}\tilde{A_0} \, \nonumber \\
&+&\,(\,a_3\,x^2\,+\,e\,x^2\,+\,e\,\alpha\,x\,y\,a_4%
\,y^2\,)\,\tilde{{\phi}^{\prime}}\tilde{A_1}  \nonumber \\
&+&\,[\,e\,\alpha\,y^2\,+\,(\,a_3\,+\,e\,+\,a_4\,)\,x\,y\,]\,\tilde{\dot{%
\phi}}\tilde{A_1} \, \nonumber \\
&+&\,[\,e\,\alpha\,x^2\,+\,(\,a_3\,+\,e\,+\,a_4\,)\,x\,y%
\,]\,\tilde{\phi^{\prime}}\tilde{A_0}  \nonumber \\
&+&\,(\,a_5\,y^2\,+\,a_7\,x\,y\,+\,a_6\,x^2\,)\,\tilde{A_0}^2\, \nonumber \\
&+&\,
(\,a_5\,x^2\,+\,a_7\,x\,y\,+\,a_6\,y^2\,)\,\tilde{A_1}^2  \nonumber \\
&+&\,[\,(\,a_7\,(\,y^2\,+\,x^2\,)\,+\,2\,(\,a_5\,+\,a_6)\,x\,y\,]\,\tilde{%
A_0}\,\tilde{A_1}\;\;,
\end{eqnarray}
where $x=sinh\,\theta$ and $y=cosh\,\theta$.

We can notice the appearance of a $\tilde{\dot{\phi}}\,\tilde{\phi^{\prime}}$
term. It does not exist in the action (\ref{final}). So, it has to
disappear. Then, we must have $a_1=-a_2$. Hence, 
\begin{equation}
{\frac{1}{4}}\,(\,\alpha^2\,+\,1\,)\,=\,\frac{\alpha^2}{\alpha^2+1}\;\;,
\end{equation}
and the solution is 
\begin{equation}
\alpha\,=\,\pm\,1\;\;.
\end{equation}
Substituting these values in (\ref{coeficientes}) we can easily see that we
can not reproduce the action (\ref{final}). So, the relativistic invariance,
in the soldering procedure, has been broken. We have now a case of
destructive interference of relativistic invariance.

\section{Conclusions}

In this work we have proposed a generalized gauged Siegel model (master action), which
can represent some gauged actions depending on the choice of the parameters.
We have promoted the fusion of two GGSM of opposite chiralities and obtained
a soldered action. The application of this action to several gauged models
already present in the literature showed new results, which can never be
obtained by a naive addition of the classical Lagrangians.

Using the Lorentz rotation to test the relativistic invariance of this master action we have fixed one of the parameters, showing that, to keep the
equivalence between the constraints in the two inertial reference systems, 
only one of the derivatives must be gauged. This is a new result about the 
issue of the chiral bosons coupled to gauge fields.

We have used the soldering formalism also to
study the action developed in \cite{dd1}. In a first step of the procedure,
we have developed a soldered action, which brings both chiralities
together. In order to keep the manifest covariance of this action, we have
demonstrated that it is not possible to find the parameter such that we have
a covariant theory showing a destructive interference of covariances. 
Interestingly we have found that the gauge field massive
term have not interfered in the process. Hence, we have looked for a value
that maintain the Lorentz invariance of the constraints. The result confirms
the one encountered in each chirality separately.

\section{Acknowledgments}

The authors would like to thank C. Wotzasek and D. Dalmazi for usefull conversations.
EMCA is financially supported by Funda\c{c}\~ao de Amparo \`a Pesquisa do Estado de S\~ao Paulo (FAPESP). 
This work is partially supported by Conselho Nacional de Pesquisa e Desenvolvimento (CNPq). FAPESP and CNPq are brazilian research agencies.

\end{document}